\documentclass[11pt]{article}

\usepackage{graphics}
\usepackage{graphicx}
\usepackage{amsthm}
\usepackage{algorithmic}
\usepackage{algorithm}
\usepackage{url}
\usepackage{subfigure}
\theoremstyle{definition}
\newtheorem{definition}{Definition}[section]
 \renewcommand{\algorithmicrequire}{\textbf{Input:}}
 \renewcommand{\algorithmicensure}{\textbf{Output:}}

\title{Extracting Dense and Connected Subgraphs in Dual Networks by Network Alignment.}

\author{Pietro Hiram Guzzi, Emanuel Salerno, Giuseppe Tradigo and Pierangelo Veltri}

\begin{document}

\maketitle

\begin{abstract}
The use of network based approaches to model and analyse large datasets is currently a growing research field. For instance in biology and medicine, networks are used to model interactions among biological molecules as well as relations among patients. Similarly, data coming from social networks can be trivially modelled by using graphs. More recently, the use of dual networks gained the attention of researchers. A dual network model uses a pair of graphs to model a scenario in which one of the two graphs is usually unweighted (a network representing physical associations among nodes) while the other one is edge-weighted (a network representing conceptual  associations among nodes). In this paper we focus on the  problem of finding the Densest Connected sub-graph (DCS) having the largest density in the conceptual network which is also connected in the physical network. The problem is relevant but also computationally hard, therefore the need for introducing of novel algorithms arises. 
We formalise the problem and then we map DCS into a graph alignment problem. Then we propose a possible solution. A set of experiments is also presented to support our approach.

\end{abstract}
\section{Introduction}

The use of network-based models to analyse data is currently growing in many research fields. For instance, in biology and medicine many approaches are based on the modelling and analysis of data using graphs \cite{Cannataro2010,di2015integrated}. Data extracted from social networks can also be  modelled using graphs and their analysis may reveal relevant information \cite{sapountzi2018social} .

Usually, many analysis approaches are based on a single network, used both to model data and to extract global and local parameters of the network as well as to identify community-related structures \cite{Clark:2014ke,Fazle2015}. In biology, community based structure are usually related to the identification of groups of related genes or proteins and their related molecular mechanisms. Diversely, in social networks, the existence of community-based structures usually indicates the presence of related users \cite{liu2018d}.

\begin{figure*}[ht]
 \label{fig:dualnetwork}
 \centering
 \includegraphics[width=0.6\textwidth]{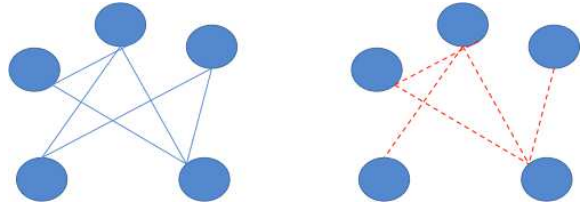}
   \caption{A dual network example. Figure shows a dual network. Graph on the left (with solid edges) represents the conceptual network network, while the other one (with red dashed edges) represents the physical network (for sack of simplicity we omitted the weight of edges on conceptual network).}
 \end{figure*}

Recently, more complex models have also been introduced. The use of a pair of graphs representing two different views of the same scenario has been introduced to detect hidden knowledge missed by simple models \cite{Wu:2016tx}. Among the others, we here focus on the so-called \textbf{the dual network model} (or dual networks). This model is based on the use of two graphs with the same vertex set and two different edge sets. One graph is unweighted and referred to as physical graph. The other graph, called conceptual graph, is edge-weighted, as depicted in Figure \ref{fig:dualnetwork}. Therefore the use of dual networks finds a natural application whenever it is needed to model two kind of relations among the same set of nodes (i.e. physical and conceptual interactions).

For instance, Phillips et al. used dual networks to analyse interactions among genetic variants \cite{Phillips:2008dm}, while Tornow et al., use dual network to analyse expression data and their functional relations \cite{Tornow:2003kc}. In such a scenario networks representing the co-expression of genes (functional networks) may be jointly analysed with other one presenting known interactions among proteins. The integration of data may help to find relations among gene co-expression and known interactions. Ulitsky et al., \cite{Ulitsky:2007iz} use a graph representing genetic interactions, i.e. a graph whose nodes are genes and edges represent the association of two genetic perturbations affecting the phenotype (\textit{genetic network}), and a graph representing physical interactions among genes (\textit{physical network}) \cite{Ulitsky:2007iz}. 

Among the others, one interesting problem that arises in dual network analysis is finding the Densest Connected Subgraph (DCS). Formally, given two input graphs $G_{p}=(V,E_{p})$ (undirected and edge-weighted), and $G_{c}=(V,E_{c})$ (undirected and unweighted), the problem consists in finding a subset of nodes $ I_{s}$ that induces a densest community in $G_{c}$ and a connected subgraph in $G_{p}$.  As proved in \cite{Wu:2016tx} the DCS problem is NP-hard in its general formulation since it may be reduced from the set cover problem \cite{Karp:2009ko}, therefore there is the need for introducing novel heuristics able to solve.

While finding a densest graph in a single network has been resolved by many approaches employing different heuristics, finding a DCS in dual networks is still a challenging problem. Wu et al., \cite{Wu:2016tx} propose two heuristics based on pruning for solving DCS problem, while we here propose a different approach. We model the problem as a local network alignment problem and we propose a novel algorithm to solve it: DN-Aligner. DN-Aligner uses a merge-and-mine approach following some previous works \cite{mina2014improving,Guzzi2012,PietroHiramGuzzi:2017bn}. It receives as input the pair of networks merging these in a single weighted \textbf{alignment graph}. Each node of the input graph represent a pair of corresponding nodes of the input ones. Each weighted edge of this graph is added using a scoring model. In this way each sub-graph of this graph represent a connected sub-graphs of the input ones. The weights of the edges are derived from the input conceptual network. Finally, we extract  densest sub-graphs of this graph by using the Charikar algorithm \cite{charikar2000greedy}. Such densest sub-graphs represent connected graph in physical network therefore they are solutions of the initial problem. 

With respect to Wu et al., our approach is more flexible, since it enables to define the input correspondence among nodes, while the approach of Wu et al., is based on the correspondence of nodes with the same name, therefoure our approach may be easily extended when node sets are not the same and our approach may be also find other kind of communities (e.g. by using a different algorithm for mining alignment graph as we present in the follows).

 \label{fig:dcsasgraph}

We provide an implementation of our algorithm and we show the effectiveness of our approach through three case studies on social networks data, on biological networks and on a co-authorship networks. Results confirm the effectiveness of our approach.

The paper is structured as follows: Section \ref{sec:related} discusses main related works, ,Section \ref{sec:dnalgorithm} presents our algorithm; Section \ref{sec:casestudies} discusses the case studies; finally Section \ref{sec:conclusion} concludes the paper.

\section{Related Work}
\label{sec:related}

Dual Networks are present in many real-life applications in which the use of a single network is unable to describe two different kinds of interactions among a set of nodes. These applications span a large of fields as introduced before: from bioinformatics to social networks. Finding a densest connected sub-graph in a dual network is different from discovery co-dense sub-graphs  or coherent
dense sub-graphs \cite{kelley2005systematic,hu2005mining,pei2005mining}.In all these approaches the algorithm manage multiple networks of the same type while in a dual network the physical and conceptual network convey different information and cannot be treated in the same way.

Moreover detecting dense components of a graph is one of the most challenging problems in graph analysis \cite{lee2010survey,khuller2009finding}. Recently, it has found many important applications in social network analysis \cite{parthasarathy2011community,ma2017detection}  as well as in bioinformatics \cite{hu2005mining}. The problem is based on the definition of \textit{density} for a graph, and literature contains many definitions of it that have been applied in different context. One of the first definition of dense sub-graph is a fully connected sub-graph, i.e. a clique. However the determination of a maximal clique, also referred to as the maximum clique problem, is NP-Hard \cite{hastad1996clique}, and it is difficult to approximate \cite{bomze1999maximum}. Moreover in real networks some edges are missing therefore the use of cliques may miss some important information.

Consequently, many definition of sub-graphs that are not fully interconnected have been introduced. Given an undirected and unweighted graph $G=(V,E)$, where $\|V\|=n$ is the set of nodes and $\|E\|=m$ is the set of edges, and a sub-graph $S=V_s,E_s$ of $G$, where $V_s \in V$, and $E_s \in E$. A common definition of density of a graph is given by the ratio of the existing edges to the maximum number of possible edges $d=\frac{\|2*E\|}{\|V\|*\|V-1\|}$. This definition may be easily extended for directed and weighted graphs. In particular for weighted graph the previous definition may be adapted by considering the weighted sum of edges and their weights. 

A \textit{densest sub-graph} is then a sub-graph with maximum density and the  densest-sub-graph problem is to find a sub-graph with maximum density. The problem may be solved in polynomial time by an algorithm developed by Goldberg based on maximum-flow approach \cite{goldberg1984finding}. Asashiro et al. \cite{asahiro2000greedily} proposed a greedy algorithm based on the strategy of deleting the node with minimum degree.
Recently, Wu et al., \cite{Wu:2016tx}, proposed an algorithm for finding densest connected sub-graph in a dual network. The approach is based on a two-step strategy. In the first step the algorithm prunes the dual network without eliminating the optimal solution. In the second step two greedy approaches are developed to build a search strategy for finding the DCS. Briefly, the first approach finds the densest sub-graph in the conceptual network first, and then it is refined to guarantee that it is connected in the physical network. The second approach maintain the sub-graph connected in the physical network while deleting low-degree nodes in the conceptual network. Authors also propose a possible solution for finding the DCS with fixed number of nodes and for maintaining a set  of input seed nodes in the identified sub-graph. 

\section{Formulation of the DCS Problem}
\label{sec:formulation}

We adopt classic formulation of DCS problem \cite{Wu:2016tx} and we summarise main notation used in this paper in table \ref{tab:formulation}.

A dual network comprises two networks sharing the same node set. One network, called physical network, has unweighted edges. A second network, called conceptual network, has weighted edges. Edge sets are in general different in the two networks.

\begin{definition}{Dual Network}
A dual network $G=(V,E_p,E_c)$ is a pair of networks: a conceptual weighted network $G_c(V,E_c)$ and a physical unweighted network $G_p(V,E_p)$.
\end{definition}

After introducing dual networks, we should give the definition of density we use in the rest of the paper. The density of an unweighted graph is given by the ration of actual number of the edges and the number of nodes.

\begin{definition}{Density of unweighted graph}\\
Given an unweighted graph $G(V,E)$ the density of the graph is defined as the number of edges and the number of nodes. $\rho=\frac{|E(V)|}{|V|}$
\end{definition}

The definition may be easily extended to weighted graphs as described in literature \cite{Wu:2016tx} by considering the sum of the weights of the edges of a node, also known as the $vol$ of a node.

\begin{definition}{Vol}\\
$vol(v)$=$\sum_{(v,w)\in E}weight(v,w)$
\end{definition}

Therefore, the density of a graph may be calculated as the average of the edge weights.

\begin{definition}{Density of a weighted network}\\
$\rho(G)=\frac{\sum_{v \in V}(vol(v)}{|V|}$
\end{definition}

Then, given a dual network we may consider the subgraphs $G_{pi}$ and $G_{ci}$ induced in the two networks by the same node set $I \subset V$. A densest common subgraph $DCS$ is a subset of nodes $I_s$ such that the density of the induced conceptual network is maximised and the induced physical network is connected.

\begin{definition}{Densest Common Subgraph}.\\ Given dual networks $G(V,E_c,E_p)$, the densest connected subgraph is a subset of nodes $I_s \subset V$ such that $G_{pI_s}$ is connected and the density of $G_{cI_s}$ is maximised.
\end{definition}{}

\begin{table}[ht]
 \caption{Main Symbols used in this work}
    \centering
    \begin{tabular}{|p{4cm}|p{6cm}|}\hline
     Symbol & Definition  \\ \hline
    $G=(V,E)$& Graph G with node set $V$ and edge set $E$\\ \hline
    $G=(V,E_p,E_c)$& a dual network made by a conceptual network $G_c(V,E_c)$ and a physical network $G_p(V,E_p)$\\ \hline
    S $\subset$ V & a subset of nodes\\ \hline
    $vol(v), v \in V$ & the sum of the weights of the edges incident to the node $v$ \\   \hline  
    $ \rho(V)=\frac{\sum_{\forall v \in V    }(vol(v))}{|V|}$ & density of a graph G defined as the average vol of the nodes \\ \hline    
    \end{tabular}
   
    \label{tab:formulation}
\end{table}{}

\section{The DN-Algorithm}
\label{sec:dnalgorithm}
The algorithm is based on the following steps as depicted in the following algorithm  \ref{algo:dcs}: \begin{itemize}
    \item Integration of the input networks into a single alignment graph;
    \item Analysis of the alignment graph using the adapted Charikar algorithm.
\end{itemize}

\begin{algorithm}
\label{algo:dcs}
\caption{DN-Aligner Algorithm}
 \begin{algorithmic}[1]
\REQUIRE: A Conceptual Network  $G_1=(W,E)$, and a Physical Network $G_2=(V,E)$,
\REQUIRE: A Correspondence File F among nodes
\REQUIRE: A distance treshold $\delta$ (optional)
\ENSURE: DCS
\STATE: AL $\leftarrow$ BuildAlignmentGraph($G_1$,$G_2$,$\delta$,F)
\STATE: DCS $\leftarrow$ Analyse(AL)
\RETURN DCS
   \end{algorithmic}
\end{algorithm}

In the first step the two networks are merged together in a single alignment graph. The algorithm has two other parameters: a file that stores the correspondence among nodes, i.e. how to merge together correspondent nodes, a distance threshold $\delta$ that represent the maximum threshold of distance that two nodes may have in the physical network (the parameter is optional and it is used to prune the possible solutions). 

In the first step the algorithm merges the input network in a single weighted \textbf{alignment graph}. Each node of the input graph represent a pair of corresponding nodes of the input ones. Each  weighted edge of this graph is added using a \textit{match-mismatch-gap} model. In this way each connected sub-graph of this graph represent a pair of connected sub-graphs of the input ones. Weight of the edges are derived from the input conceptual networks without modifying them. Finally we extract  densest sub-graphs of this graph by using the Charikar algorithm \cite{charikar2000greedy}. Such densest sub-graphs represent connected graph in physical network therefore they are solutions of the initial problem.

\subsection{Creation of the Alignment Graph}

The first step of our algorithm is based on a previous work on graph alignment by our group \cite{mina2014improving}, which has been modified for the purposes of this work.

\begin{figure}[htp!]
\centering
\begin{subfigure}
\centering
\includegraphics[width=0.5\textwidth]{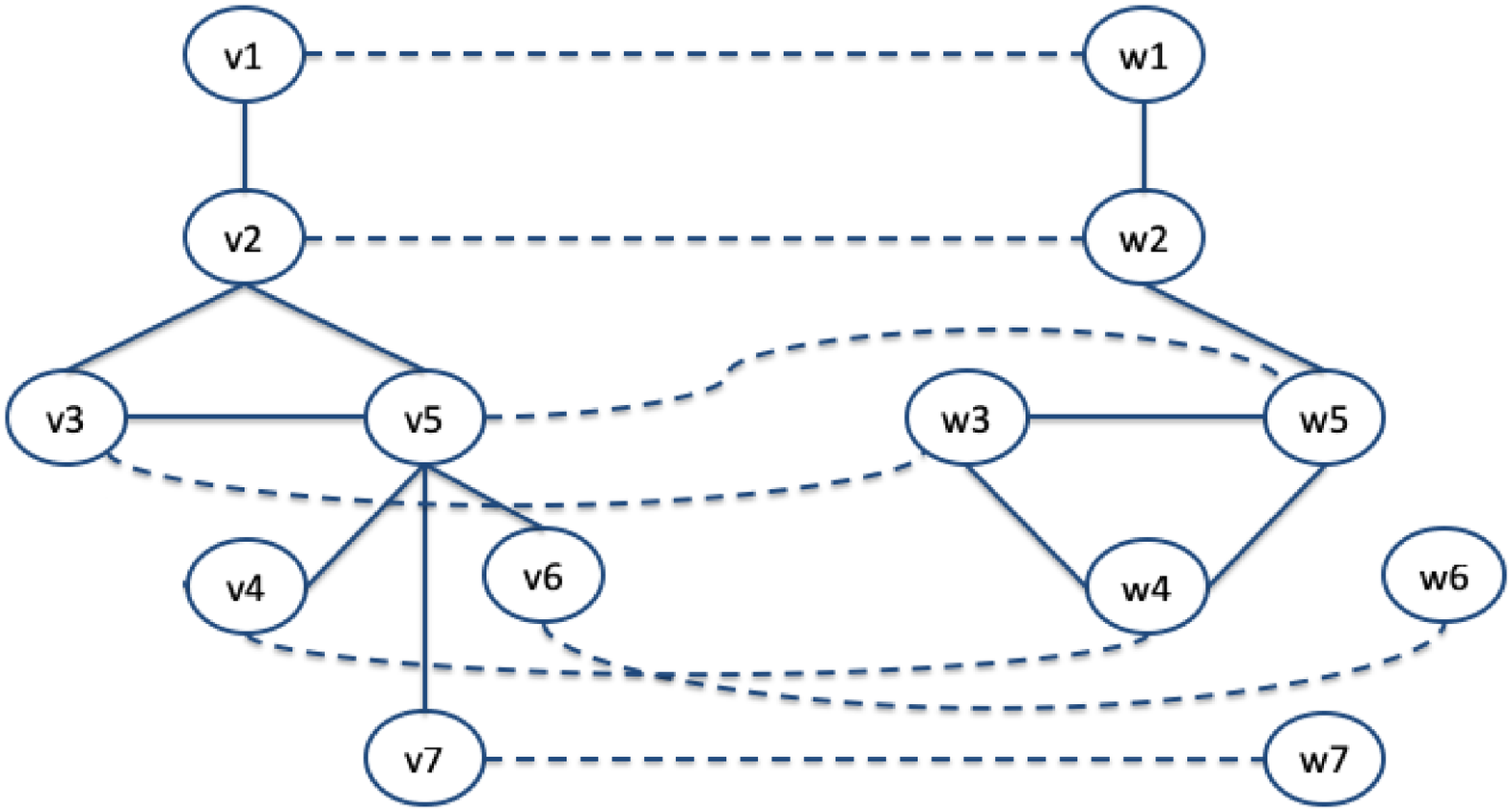}
    \caption{Alignment Example: the Algorithm receives as input two networks and a set of similarity relationship among nodes of the networks (dashed lines).}
    \label{fig:allesempio2}
\end{subfigure}
\begin{subfigure}
\centering
\includegraphics[width=0.5\textwidth]{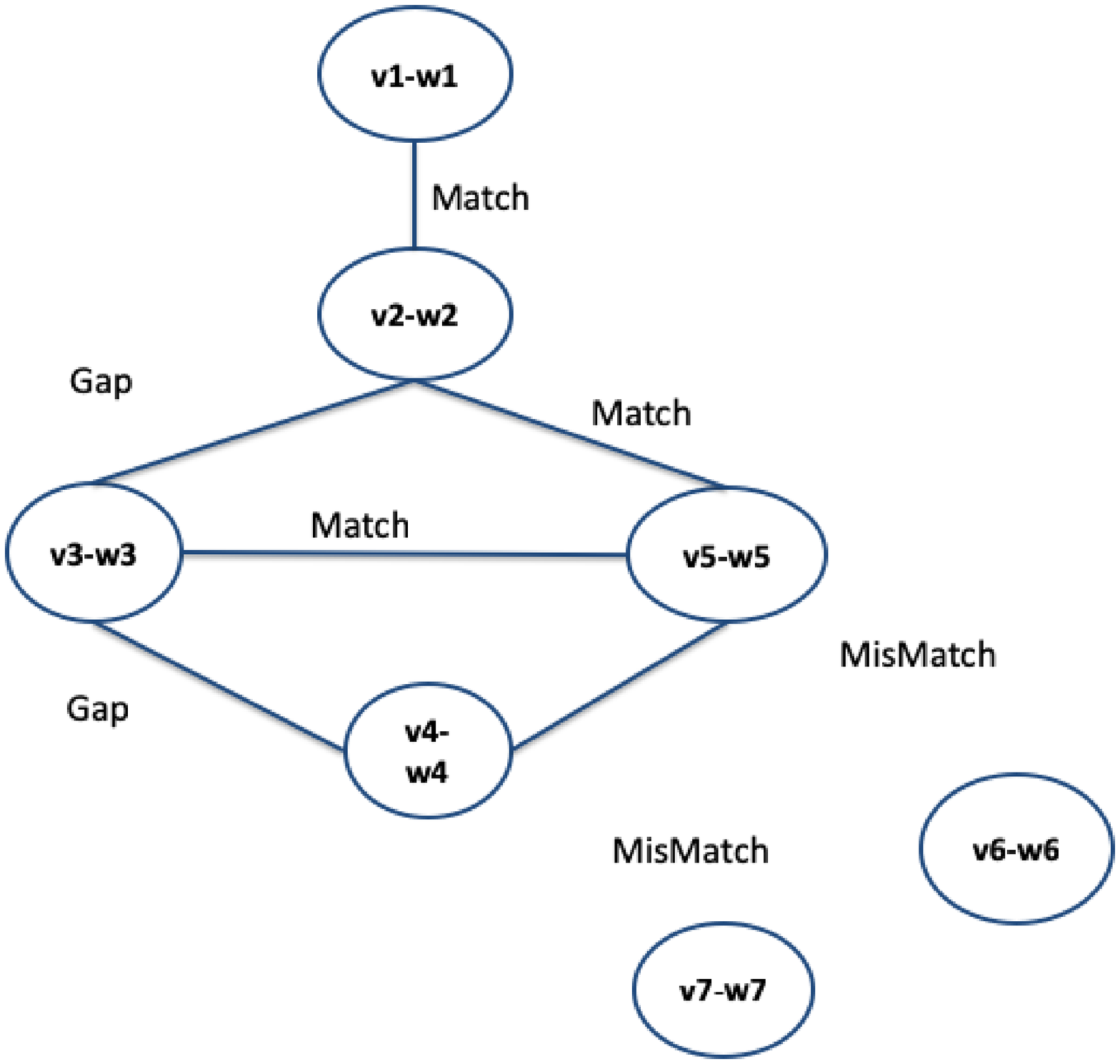}
    \caption{First, the algorithm builds the nodes of the heterogeneous alignment graph. The edges are then added according to the analysis of input networks.}
\end{subfigure}
\label{fig:allesempio2bis}
\end{figure}

We explain the building of the alignment graph through an example. Let us consider two input graphs: a weighted graph $G_1=(W,E)$, and an unweighted graph  $G_2=(V,E)$, as depicted in Figure \ref{fig:allesempio2}. The proposed algorithm builds the alignment graph by considering both the input graphs and a set of relations of similarity among nodes used as the seed.  For the sake of simplicity, networks have the same number of nodes. Figure \ref{fig:allesempio2} shows these relationships as dashed lines connecting the nodes of the two graphs. 

First, the algorithm builds a new node, defined as \textbf{composite node}, for each pair of nodes that are in a relationship. Each node of the alignment graph represents a pair of correspondent nodes. After this step, the algorithm adds the edges among nodes by examining the two input graphs. An edge between two nodes is inserted whenever the corresponding nodes are connected in both the input networks. For instance, both $(v1,v2)$ and $(w1,w2)$ are both connected in input networks, hence the alignment graph will contain $(v1-w1)$ and $(v2-w2)$ nodes. This condition represents a  \textbf{Match}. Therefore an edge is inserted between $(v1-w1)$ and $(v2-w2)$ and the weight of this edge is the weight of the edge $w1-w2$. Let us consider nodes $(v3-w3)$ and $(v4-w4)$ of the alignment graph. Nodes $w3$ and $v4$ are adjacent while $v3$ and $v4$ are connected but not adjacent and the distance is below a given a threshold of distance $\delta$, (for instance $4$). In this case an edge will be inserted between nodes $(v3-w3)$ and $(v4-w4)$ of the adjacent graph, while the weight of the edge is the average of the weights of the edges of the path linking $w3$ to $w4$. After the analysis of all node pairs, the final alignment graph is built, as represented in Figure \ref{fig:allesempio2bis}. The analysis of this graph is the second step of our algorithm.

Formally, as explained in the following algorithm, in the step 1 the procedure \texttt{BuildAlignmentGraph} receives as input the two networks, a set of relations among their nodes stored in the similarity file $F$ and a threshold $\delta$ and it produces as output a weighted alignment graph Al=$G_{al},E_{al}$ . The similarity file contains a set of node pairs, and each pair contains a node of each network. The similarity file contains one-to-one relations, i.e. each node of a network is linked to a single node of the other network. First, the algorithms build the node of the alignemnt graph in a trivial way. The procedure scans the similarity file and for each pair of nodes it builds a node of the alignment graph. Then it considers all pairs of nodes of Al. Given two nodes of the alignment graph $v_{al,1}=(v_{1},w_{1})$ and $v_{al,2}=(v_{2},w_{2})$, it adds a corresponding edge between them when the input nodes are adjacent in the two input networks. In this case the weight of the edge in Al is the weight of the corresponding edge in the conceptual network. 

Given two nodes of the alignment graph $v_{al,1}=(v_{1},w_{1})$ and $v_{al,2}=(v_{2},w_{2})$, a \textbf{gap} is proved when the input nodes are adjacent only in the conceptual network network and they are at distance lower than  $\Delta$ (gap threshold) in the physical network. In this case an edge will be inserted in to Al and the weight is the average weight of the edges of the shortest path connecting them. Conversely when they are at distance greated than $\Delta$, no edges are inserted into the alignment graph. When $\Delta$ is set to $\infty$, this means that an edge is inserted whenever the nodes are connected into the physical network.

\begin{algorithm}
 \caption{Step 1: Building the Weighted Alignment Graph.}
 \label{algo:AlignmentGraph}
 \begin{algorithmic}[1]
 \renewcommand{\algorithmicrequire}{\textbf{Input:}}
 \renewcommand{\algorithmicensure}{\textbf{Output:}}
 \REQUIRE ($G_1$,$G_2$,$\delta$,F)
 \ENSURE  Al=$G_{al},E_{al}$ (weighted Alignment Graph
 \\ \textit{Initialisation} :
  \STATE Building of Nodes
 \\ \textit{LOOP Process: Scan of F file}
  \FOR {pair contained in F}
  \STATE add Node to $G_{al}$
  \ENDFOR
 \STATE Building of Edges
 \\ \textit{For Each Node $\in$ $G_{al}$}
  \FOR {Node $\in$ $G_{al}$  }
  \STATE  $E_{al}$ $\leftarrow$ Analyse($G_1$,$G_2$)
 \ENDFOR
  \RETURN $AL$ 
 \end{algorithmic} 
\end{algorithm}

\subsection{Densest Connected Graph Extraction: The Adapted Charikar Algorithm.}

The Charikar algorithm produces a densest sub-graph $S$ of given graph $G$ by using a greedy approximation. The algorithm has been developed initially for unweighted graphs. The idea behind the algorithm  is that the elimination of low degree vertices in a unweighted graph may produce a sub graph $S$
having the desired properties. The algorithm starts by considering the whole graph $G$. For each iteration it identifies the minimum degree vertex $v_{min} \in G $ and it removes $v_{min}$ from $G$. The algorithm stops when all the vertices have been removed from  $G$. The sub-graph with maximum density is built and returned as output during the iterations. It has been proved that it represents a 2 approximation for the problem and that the algorithm run in $\mathcal{O}(n+m)$, where $n$ and $m$ are the nodes and the edges of the input graph. The algorithm may be easily extended to weighted graph by considering the weighted sum of the degree and weights \cite{charikar2000greedy}. In our approach we used the extension of the Charikar algorithm for weighted graph and we provided a Python implementation. Our algorithm differs from the Charikar original definition onky for the definition of density as follows.

Let $G=(V,E)$ an undirected graph with weighted edges and $S \subset V$ a sub-graph. Each node (v) as a set of incident edges $(E(v))$ and each edge as an associated weight $w$.
We define as $vol(v), v \in V$ the sum of the weights of the edges incident to the node $v$,  $vol(v)=\sum(E(v)$. We define as density of G $\rho(V)$ the ratio among the $vol(v)$ and the number of nodes of $G$: 
\begin{equation}
\rho(V)=\frac{\sum_{\forall v \in V    }(E(v))}{|V|},
\end{equation}

\section{Case Studies}
\label{sec:casestudies}

This section presents some case studies on a social network, on a co-authorship network and on a biological network. As proof of concept we present three case studies on three different networks: (i) a social network, (ii) a biological network and (iii) a co-authorship network. In each study we extract the densest connected graph.

All the programs are written in Python Programming Language and are available for download at \url{https://github.com/hguzzi/DualNetworkAligner}. All experiments are performed on a server with 16Gb Memory, Ubuntu OS and Intel Core i5 CPU.


\subsection{Experiments on Social Networks: The GoWalla dataset.}

\begin{figure}
    \centering
    \includegraphics[width=0.7\textwidth]{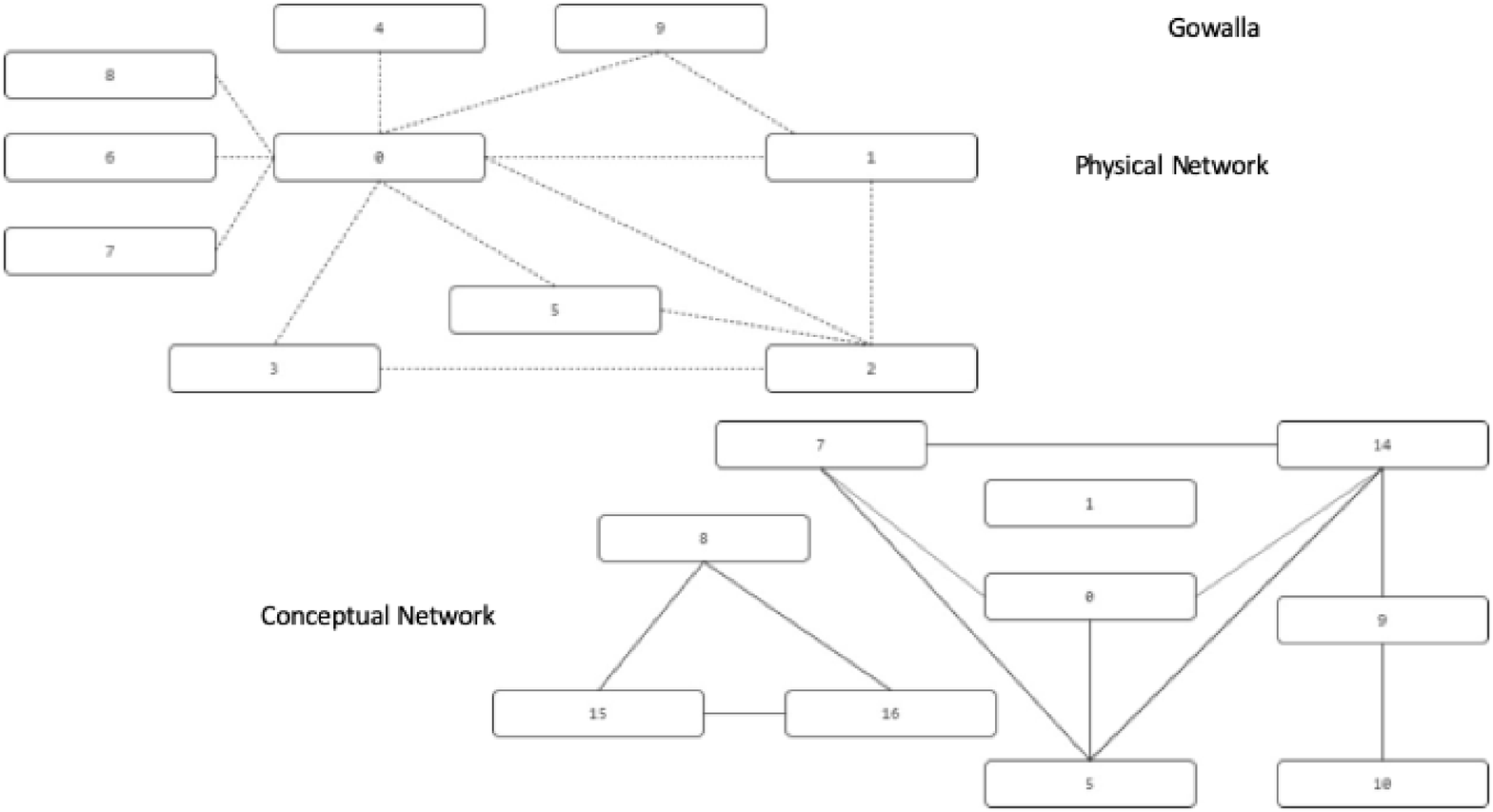}
    \caption{Dual Network representing GO-Walla Users.}
    \label{fig:gowalla}
\end{figure}

Gowalla is a social network where users share their locations (expressed as GPS coordinates) by checking-in into the web-site \cite{cho2011friendship}. We downloaded data contained in SNAP datasets collection \cite{snapnets}. The whole network is undirected and it consists of   196,591 nodes and 950,327 edges. Each node represents a user and each edge link two friends into the network. In order to obtain a dual network we considered two possible network starting from these data. We realised a  a physical network represents the friendship network. Figure \ref{fig:gowalla} depicts an extract of the dual network.

Therefore each user of GoWalla is represented by a node, while an edge represents a friendship relation derived from data. Since each user has associated information about  position we calculated the distances among the users expressed as distance among check-ins. In case of multiple check-ins we considered the average of all the check-ins.  Then we normalised all the distances by considering the maximum distance among all the users. Therefore nodes representing users that may be considered close will be connected by edges having a weight close to one, while a weight close to zero will represent user whose positions are not close. It should be noted that two users that are geographically near may be not friend, and that two friend may be far geographically. A \textit{densest common sub-graph} in this case represents a set of users that are very close geographically and that are connected among them in a friendship network. The analysis of the conceptual network alone may miss all the information about the friendships. The extracted DCS contains 2442 nodes and 149530 edges. This community  represents a set of users that are friends and that are close from a geographical point of view.


\subsection{Experiments on Co-Authorship Network}

We evaluate our approach in a dual network representing authors and the similarity of the activity of their research. We use the DBLP dataset \footnote{The dblp team: dblp computer science bibliography. Monthly snapshot release of November 2019. https://dblp.org/xml/release/dblp-2019-11-01.xml.gz}. We considered published papers in five bioinformatics conferences: BCB, BIBM ISMB, RECOMB and EMBC. For each conference we extracted all the information about papers and authors. The dataset contains 20,563 authors.

The physical network represents co-authorship relations, therefore each node represents an author and an edge links two author that have co-authored a paper. The conceptual network model the research interest similarity among authors and it is constructed by analysing the similarity of the paper titles. We considered the Jaccard Index to compute the research interest similarity.  We obtained two graphs having 20,563 nodes, the physical network has 58536 edges while the conceptual network has 200530 nodes.  

\begin{figure}[ht]
    \centering
    \includegraphics[width=0.6\textwidth]{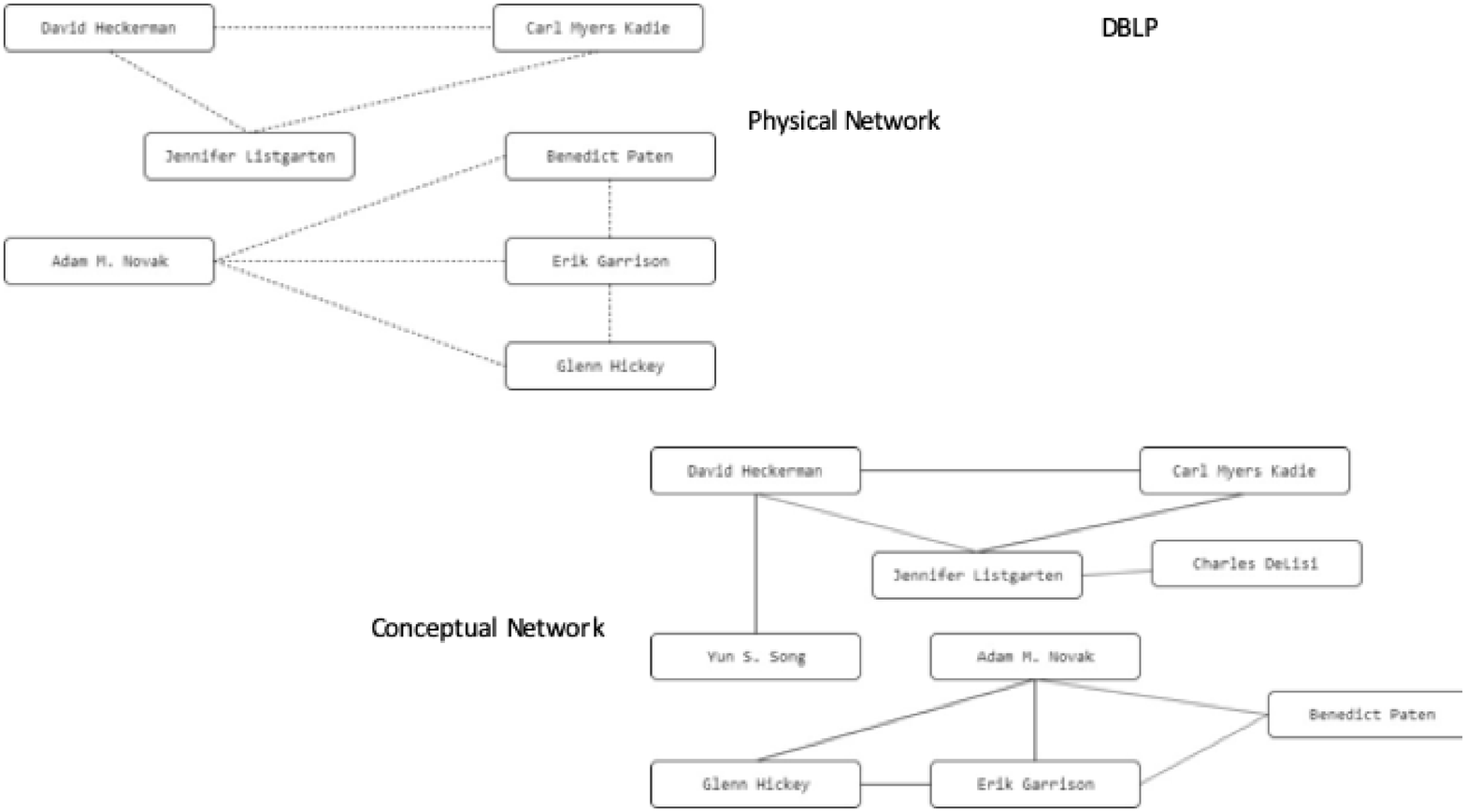}
    \caption{DCSual Network representing Co-Authorship and Similar Interests.}
    \label{fig:dblpnetwork}
\end{figure}

It should be evidenced that a dense sub-graph in the conceptual network represent a set of authors that have a great research interest similarity that may be not collaborators considering the co-authorship networks. Therefore the analysis of the only conceptual network may miss information about the chain of collaborations evidencing the need for the use of dual networks. The found DCS has 573 nodes and 95823 edges, Figure \ref{fig:dblpnetwork} depicts an extract of the found DCS. The DCS contains co-authors that share common research interests.

We also extracted a denese subgraph only in the conceptual network and we derived the induced subgraph in the co-author network. We obtained a graph with  1073 nodes and 198746 edges. We found that this graph is not connected in the physical network thus the analysis of only a network is missing many important information.

\subsection{Experiments on biological data: analysis of protein interactions.}

We considered data from the STRING database \cite{stringdatabase}. This database contains data about proteins and their interactions. Each node represents a protein and each edge takes into account the reliability of the interaction between two proteins with a value in the interval $(0-1)$. Therefore, we obtained two networks:
\begin{itemize}
\item a conceptual network, which represents the strength of associations among proteins;
\item a physical network, which stores the binary interactions among proteins.
\end{itemize}
We obtained two networks having 19.354 nodes and 5.879.727 edges. We ran our algorithm and it resulted in a DCS having 756 nodes and 154.142 edges. We performed a biological interpretation of the results by using a functional enrichment algorithm provided by the DAVID software \cite{da2007david}. Main enriched functions of the DCS are:

\begin{itemize}
    \item GO:0006281 - DNA repair
    \item GO:0006302 - double-strand break repair
    \item GO:0070182 - DNA polymerase binding
    \item GO:0003676 - nucleic acid binding
\end{itemize}

\begin{figure}
    \centering
    \includegraphics[width=0.6\textwidth]{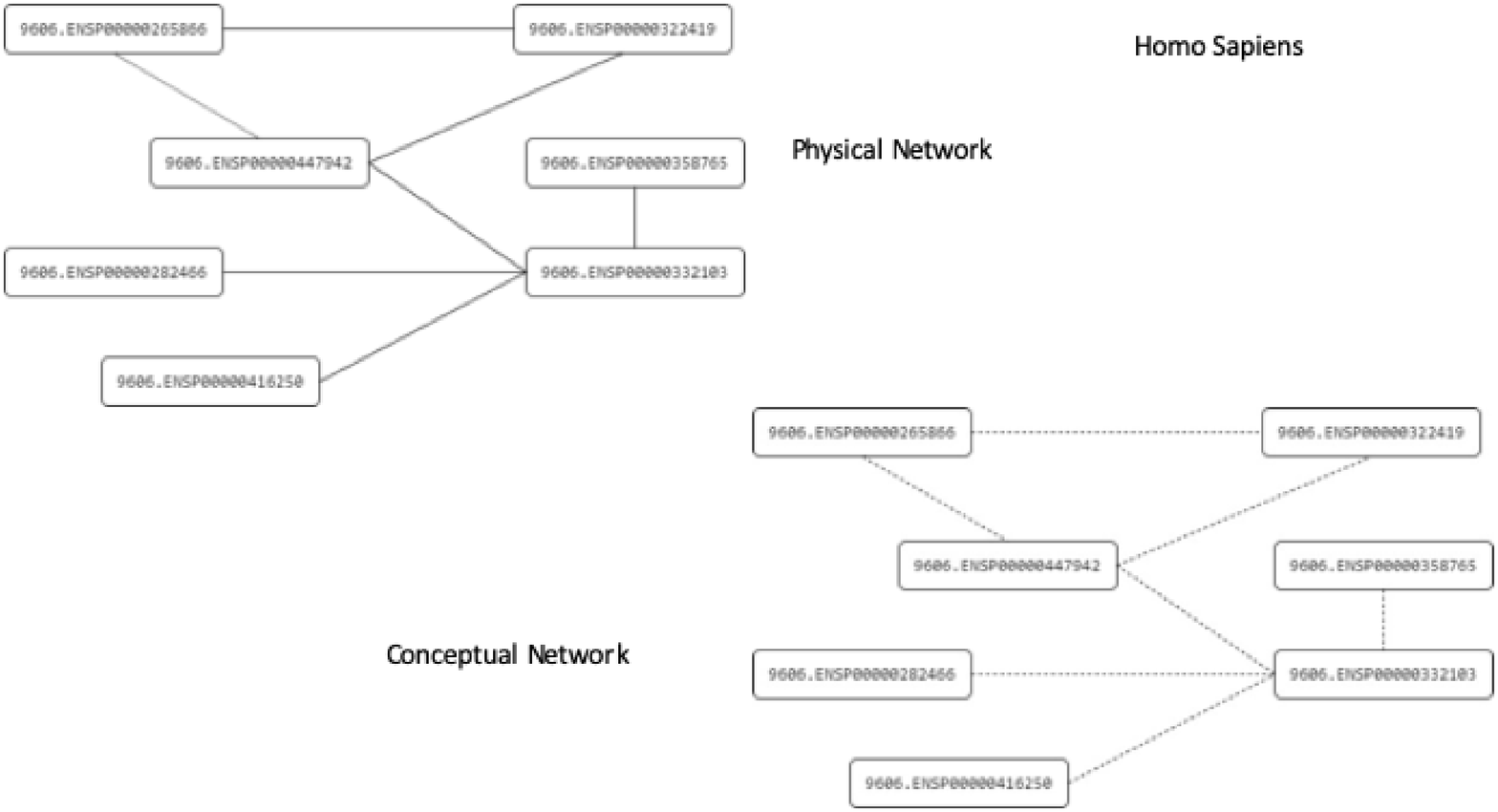}
    \caption{Dual Network representing the biological network.}
    \label{fig:homosapiens}
\end{figure}

Similarly to social networks we extracted the densest graph in conceptual network and we verified that the induced graph in the physical one is not connected.




\section{Conclusion}
\label{sec:conclusion}

In a dual network model a pair of graphs is used to model complex scenarios in which one of the two graph is unweighted (physical network) while the other is edge-weighted (conceptual network). In the present paper we presented an heuristic algorithm for obtaining the densest connected sub-graph (DCS) having the largest density in the conceptual network and being also connected in the physical network. We formalised the problem and we then mapped the DCS problem into a graph alignment problem. Finally,  we proposed a possible solution and presented a set of experiments, which demonstrate the effectiveness of our approach.

\bibliographystyle{plain}
\bibliography{Biblio.bib}

\end{document}